\documentclass[aps,prl,showpacs,twocolumn,superscriptaddress]{revtex4}
\usepackage{bm,color}
\usepackage{graphicx}
\usepackage{amsmath}
\usepackage{color}

\begin{document}
%\draft
\title {Microwave magnetochiral effect in Cu$_2$OSeO$_3$}

\author{Masahito Mochizuki}
\affiliation{Department of Physics and Mathematics, Aoyama Gakuin University, Sagamihara, Kanagawa 229-8558, Japan}
\affiliation{PRESTO, Japan Science and Technology Agency, Kawaguchi, Saitama 332-0012, Japan}
\begin{abstract}
We theoretically find that in a multiferroic chiral magnet Cu$_2$OSeO$_3$, resonant magnetic excitations are coupled to collective oscillation of electric polarization, and thereby attain simultaneous activity to ac magnetic field and ac electric field. Because of interference between these magnetic and electric activation processes, this material hosts gigantic magnetochiral dichroism on microwaves, that is, the directional dichroism at gigahertz frequencies in Faraday geometry. The absorption intensity of microwave differs by as much as $\sim$30$\%$ depending on whether its propagation direction is parallel or antiparallel to the external magnetic field.
\end{abstract}
%%75.85.+t Magnetoelectric effects, multiferroics
%%76.50.+g FM, AM, and ferrimagnetic resonances; spin-wave resonance
%%77.80.-e Ferroelectricity and antiferroelectricity
%%78.20.Ls Magneto-optical effects
%%78.20.Bh Theory, models, and numerical simulation
%%78.70.Gq Microwave and radio-frequency interactions
\maketitle
%\sloppy \maketitle
Collective excitations of spins in magnets, so-called magnons or spin waves, can be activated not only via a direct process with ac magnetic field $\bm H^\omega$ coupled to magnetizations but also via an electric excitation by ac electric field $\bm E^\omega$ coupled to charge degrees of freedom. When the magnon or spin-wave modes have simultaneous activity to the $\bm H^\omega$ and $\bm E^\omega$ components of electromagnetic waves, interference between the two activation processes, that is, the magnetically activating and the electrically activating processes, gives rise to peculiar optical and/or microwave phenomena, so-called optical ME effect. One of the most important examples is the directional dichroism, that is, oppositely propagating electromagnetic waves exhibit different absorptions.
%where an electromagnetic wave propagating in a certain direction can well penetrate 
%the sample, whereas that propagating in the opposite direction becomes strongly 
%absorbed by the sample and cannot penetrate it.

Multiferroic materials with concurrent magnetic and ferroelectric orders~\cite{Kimura03a,Katsura05,Fiebig05,Tokura06,Tokura07,Cheong07,Eerenstein06,Kimura07} provide an opportunity to realize the electric-dipole active magnons (so-called electromagnons)~\cite{Smolenski82,Katsura07,Pimenov06a,Pimenov06b,Kida09}, and thus the optical ME effect via the magnetoelectric coupling~\cite{Szaller13,Miyahara13,Miyahara14}. Indeed observations of the directional dichroism have been reported for several multiferroic materials such as Ba$_2$CoGe$_2$O$_7$~\cite{Kezsmarki11,Bordacs12,Kezsmarki14}, $R$MnO$_3$ ($R=$rare-earth ions)~\cite{Takahashi12,Takahashi13}, and CuFe$_{1-x}$Ga$_x$O$_2$~\cite{Kibayashi14}, in which nontrivial spin orders induce the ferroelectric polarization via the relativistic spin-orbit interaction. In these materials, the optical ME effect is observed at the electromagnon resonance frequencies in the terahertz (THz) regime.

The directional dichroism is observed also at higher frequencies, i.e., x-ray and visible-light regimes in several polar magnets, which is caused by electron transitions among the spin-orbit multiplets~\cite{Rikken97,Rikken02,Arima08,Kubota04,Jung04,Saito08a,Saito08b}. However, observations of the effect at gigahertz (GHz) frequencies are quite limited and the effect observed so far is very tiny whose difference in absorption intensity is only $2.5\%$ at most~\cite{Okamura13}, while the directional dichroism at GHz frequencies is anticipated for application to microwave devices~\cite{Gurevich96}. This is because most of the well-known multiferroic materials based on simple spiral or antiferromagnetic spin structures with short-period modulation tend to have relatively large spin-wave gaps of several meV, which inevitably results in rather high resonance frequencies in THz regime. To achieve the microwave ME effect, long-period magnetic textures with tiny spin-wave gap should be examined, and slowly modulating magnetic structures induced by the Dzyaloshinskii-Moriya interaction (DMI) are promising for this purpose. Indeed, the nonreciprocal directional dichroism of microwaves in Voigt geometry was theoretically predicted for the DMI-induced skyrmion phase~\cite{Mochizuki12}.%%In Ref.~\ref{Mochizuki12}, we have theoretically examined the 
%%skyrmion-crystal phase in a thin-film specimen of the chiral-lattice 
%%insulator Cu$_2$OSeO$_3$.

In this Letter, we theoretically predict that electromagnon excitations in magnetically ordered phases of Cu$_2$OSeO$_3$ exhibits unprecedentedly large magnetochiral dichroism at GHz frequencies, that is, the microwave directional dichroism in Faraday geometry. In the presence of $\bm H$, the conical spin phase or the field-polarized ferrimagnetic phase emerges in the bulk samples depending on the strength of $\bm H$. When $\bm H$ is applied in a certain direction, and a microwave is irradiated parallel or antiparallel to $\bm H$, the absorption intensities for the oppositely propagating microwaves differ by as much as 30$\%$. Such a huge directional dichroism at microwave frequencies has never been observed in a single material. 
This effect is traced back to the resonantly enhanced magnetoelectric coupling, and therefore essentially distinct in microscopic mechanism from the traditional microwave non-reciprocal device based on microwave polarization, potentially leading to a unique microwave device.
%%This effect is resonantly enhanced at the eigen-frequency of the 
%%electromagnon mode ($\sim$GHz), which can be tuned by an external 
%%magnetic field. Our finding offers a concept for designing novel 
%%microwave devices such as magnetically tunable microwave isolator.

%%%%%%%%%%%%%%%%%%%%%%%%%%%%%%%%%%%%%%%%%%%%%%%%%%%%%%%%%%%%%
\begin{figure}
\includegraphics[scale=1.0]{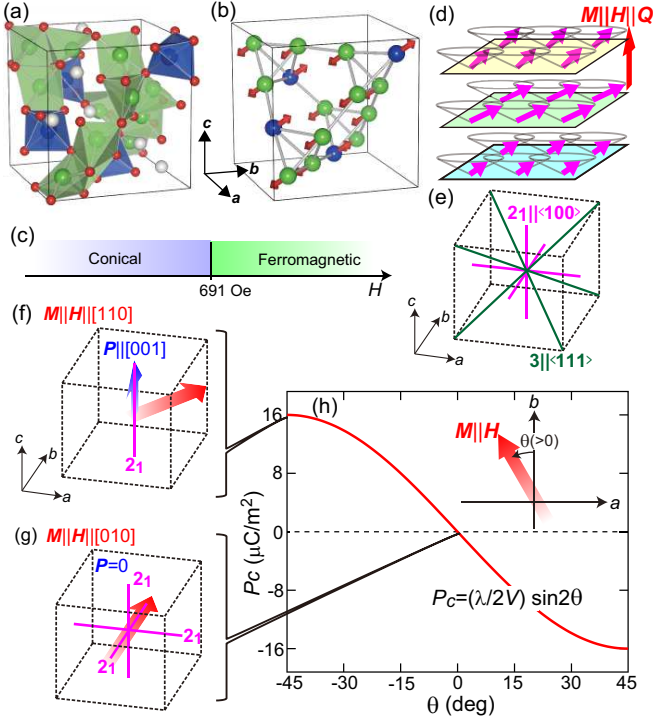}
\caption{(color online). (a) Crystal structure of Cu$_2$OSeO$_3$. (b) Magnetic structure of Cu$_2$OSeO$_3$. (c) Phase diagram of the spin model given by Eq.~(\ref{eqn:model}) with $J$=1 meV and $D/J$=0.09. (d) Schematic figure of the conical spin structure. (e) Symmetry axes in the chiral cubic P2$_1$3 crystal. (f) Ferroelectric polarization $\bm P \parallel [001]$ under $\bm H \parallel [110]$. (g) Absence of  $\bm P$ under $\bm H \parallel [010]$. (h) Calculated net polarization $P_c$ in the field-polarized ferromagnetic state under $\bm H$$\perp$$\bm c$ as a function of the angle $\theta$ between $\bm H$ and the $b$ axis.}
\label{Fig1}
\end{figure}
%%%%%%%%%%%%%%%%%%%%%%%%%%%%%%%%%%%%%%%%%%%%%%%%%%%%%%%%%%%%%
%%%%%%%%%%%%%%%%%%%%%%%%%%%%%%%%%%%%%%%%%%%%%%%%%%%%%%%%%%%%%
%%\begin{figure*}
%%\includegraphics[scale=1.0]{Fig1.eps}
%%\caption{(color online). (a) Crystal structure of Cu$_2$OSeO$_3$. (b) Magnetic structure of Cu$_2$OSeO$_3$ whose magnetic unit is a tetrahedron composed of four Cu$^{2+}$ ions ($S$=1/2). In each tetrahedron, collinear three-up and one-down spin configuration is realized below $\sim$58 K. (b) Phase diagram of the spin model given by Eq.~(\ref{eqn:model}) with $J$=1 meV and $D/J$=0.09, in which a transition from the longitudinal conical phase to the ferromagnetic phase occurs at $H$=691 Oe. (c) Schematic figure of the conical spin structure under an external magnetic field $\bm H$ which has a propagation vector $\bm Q \parallel \bm H$ and a net magnetization $\bm M \parallel \bm H$. (e) Symmetry axes in the Cu$_2$OSeO$_3$ crystal, which belongs to the cubic P2$_1$3 space group: three-fold rotation axes, 3, along $\left< 111 \right>$ and two-fold screw axes, $2_1$, along $\left< 100 \right>$. (a) Emergence of ferroelectric polarization $\bm P \parallel [001]$ is allowed under $\bm H \parallel [011]$ from the symmetry. (b) Ferroelectric polarization $\bm P$ should be absent under $\bm H \parallel [010]$ from the symmetry. (c) Calculated net polarization $P$ in the ferromagnetic state under $\bm H$$\perp$$\bm c$ as a function of the angle $\theta$ between $\bm H$ and the $a$ axis.}
%%\label{Fig1}
%%\end{figure*}
%%%%%%%%%%%%%%%%%%%%%%%%%%%%%%%%%%%%%%%%%%%%%%%%%%%%%%%%%%%%%
The crystal and magnetic structures of Cu$_2$OSeO$_3$ consist of a network of tetrahedra composed of four Cu$^{2+}$ ($S$=1/2) ions at their apexes as shown in Figs.~\ref{Fig1}(a) and (b). In each tetrahedron, three-up and one-down collinear spin configuration is realized below $T_{\rm c}$$\sim$58 K~\cite{Bos08,Belesi10}. This four-spin assembly can be regarded as a magnetic unit, and is described by a classical magnetization vector $\bm m_i$ whose norm $m$ is unity. We employ a classical Heisenberg model on a cubic lattice to describe the magnetism in a bulk specimen of Cu$_2$OSeO$_3$~\cite{Bak80,YiSD09,HanJH10}, which contains the ferromagnetic-exchange interaction and the Dzyaloshinskii-Moriya interaction among the effective magnetizations $\bm m_i$ and the Zeeman coupling to the external $\bm H$. The Hamiltonian is given by,
%%%%%%%%%%%%%%%%%%%%%%%%%%%%%%%%%%%
\begin{eqnarray}
\mathcal{H}_0&=&
-J \sum_{<i,j>} \bm m_i \cdot \bm m_j
-D \sum_{i,\hat{\bm \gamma}} 
\bm m_i \times \bm m_{i+\hat{\bm \gamma}} \cdot \hat{\bm \gamma} 
\nonumber \\
&-&g\mu_{\rm B}\mu_0 \bm H \cdot \sum_i \bm m_i,
\label{eqn:model}
\end{eqnarray}
%%%%%%%%%%%%%%%%%%%%%%%%%%%%%%%%%%%
where $g$=2, and $\hat{\bm \gamma}$ runs over $\hat{\bm a}$, $\hat{\bm b}$, and $\hat{\bm c}$ in the cubic setting. Here details of the exchange interactions within each tetrahedron are neglected and only the effective interactions among these magnetic units $\bm m_i$ are taken into account. Such a coarse graining is justified as far as low-energy excitations with long-wave-length magnetic modulations are considered. We take $J$=1 meV and $D/J$=0.09 so as to reproduce the experimentally observed $T_{\rm c}$ and periodicity ($\sim$ 50 nm) of the conical state.

Figure~\ref{Fig1}(c) shows a phase diagram of this spin model at $T$=0 as a function of the magnetic field $H$, which exhibits a phase transition between the conical phase and the ferromagnetic phase at $H$=691 Oe, in agreement with the experiments for the bulk specimen~\cite{Seki12a,Adams12}. Schematic figure of the spin structure in the conical phase is shown in Fig.~\ref{Fig1}(d), in which both the propagation vector $\bm Q$ and the net magnetization $\bm M$ are parallel to $\bm H$. Note that the spin textures considered here are slowly varying in space, and thus the coupling to the background crystal structure is significantly weak. This justifies our treatment with a spin model on the cubic lattice after the coarse graining of magnetizations.

The presence or absence of the ferroelectric polarization $\bm P$ and, if any, its direction can be known from the symmetry consideration~\cite{Seki12a,Seki12b}. The crystal structure of Cu$_2$OSeO$_3$ belongs to the chiral cubic P2$_1$3 point group, which has three-fold rotation axes, 3, along $\left< 111 \right>$, and two-fold screw axis, $2_1$, along $\left< 100 \right>$ as shown in Fig.~\ref{Fig1}(e). This crystal symmetry is not polar, and thus there exists no spontaneous $\bm P$. Although the conical and the ferromagnetic spin states are not polar, either, combination of the crystal and the magnetic symmetries renders the system polar, and allows the emergence of $\bm P$. As shown in Fig.~\ref{Fig1}(f), the emergence of $\bm P$$\parallel$[001] perpendicular to the net magnetization $\bm M$($\parallel$$\bm H$) is expected for the conical and the ferromagnetic states formed under $\bm H$$\parallel$[110] since only the $2_1$ axis along [001] remains as a symmetry axis. On the other hand, the emergence of $\bm P$ is forbidden under $\bm H$$\parallel$[010] since three $2_1$ axes survive as shown in Fig.~\ref{Fig1}(g).

Microscopically the local polarization $\bm p_i$ at the $i$th tetrahedron is given using the magnetization components $m_{ia}$, $m_{ib}$, and $m_{ic}$ as,
%%%%%%%%%%%%%%%%%%%%%%%%%%%%%%%%%%%
\begin{eqnarray}
\bm p_i=\left(p_{ia}, p_{ib}, p_{ic} \right)
= \lambda \left(m_{ib}m_{ic}, m_{ic}m_{ia}, m_{ia}m_{ib} \right).
\end{eqnarray}
%%%%%%%%%%%%%%%%%%%%%%%%%%%%%%%%%%%
The net magnetization $\bm M$ and the ferroelectric polarization $\bm P$ are calculated by sums of the local contributions as $\bm M$=$\frac{g\mu_{\rm B}}{NV}\sum_{i=1}^{N} \bm m_i$ and $\bm P$=$\frac{1}{NV}\sum_{i=1}^{N} \bm p_i$, respectively, where the index $i$ runs over the Cu-ion tetrahedra, $N$ is the number of the tetrahedra, and $V$(=1.76$\times$10$^{-28}$ m$^3$) is the volume per tetrahedron. The coupling constant $\lambda$ is evaluated as $\lambda$=$5.64\times10^{-27}$ $\mu$Cm from the experimental data~\cite{Mochizuki12}. Figure~\ref{Fig1}(h) shows calculated net polarization $P$ in the field-polarized ferromagnetic state when $\bm H$ is applied within the $c$-plane as a function of the angle $\theta$ between $\bm H$ and the $\bm b$ axis (see the inset). We find that the positive (negative) $\bm P$ emerges along [001] when $\theta$ is negative (positive).

%%%%%%%%%%%%%%%%%%%%%%%%%%%%%%%%%%%%%%%%%%%%%%%%%%%%%%%%%%%%%
\begin{figure}
\includegraphics[scale=1.0]{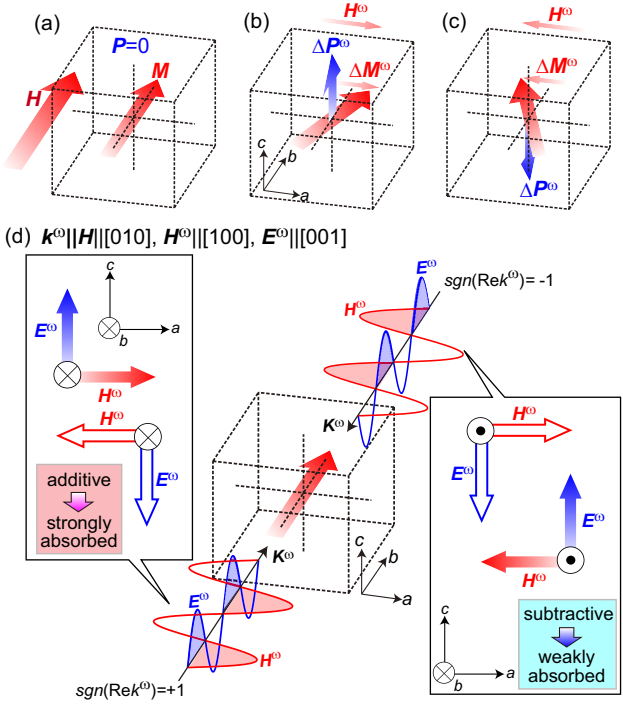}
\caption{(color online). (a)-(c) In the presence of net magnetization $\bm M$$\parallel$$\bm H$ under $\bm H$$\parallel$[010], oscillating magnetization component $\Delta \bm M^\omega$($\parallel$[100]) is accompanied by the oscillating polarization component $\Delta \bm P^\omega$($\parallel$[001]). (d) Configuration of microwave $\bm H^\omega$ and $\bm E^\omega$ components, for which the magnetochiral dichroism occurs under $\bm H$$\parallel$[010]: $\bm k^\omega$$\parallel$$\pm \bm H$, $\bm H^\omega$$\parallel$[100] and $\bm E^\omega$$\parallel$[001].}
\label{Fig2}
\end{figure}
%%%%%%%%%%%%%%%%%%%%%%%%%%%%%%%%%%%%%%%%%%%%%%%%%%%%%%%%%%%%%
According to Fig.~\ref{Fig1}(h), one realizes that oscillation of $\bm M$ ($\Delta \bm M^\omega$$\parallel$[100]) induces oscillation of $\bm P$ ($\Delta \bm P^\omega$$\parallel$[001]) when $\bm M$$\parallel$$\bm H$$\parallel$[010] [see Figs.~\ref{Fig2}(a)-(c)], and conversely $\Delta \bm P^\omega$$\parallel$[001] induces $\Delta \bm M^\omega$$\parallel$[100] via the magnetoelectric coupling. This means that both $\bm H^\omega$$\parallel$[100] and $\bm E^\omega$$\parallel$[001] components of microwave can activate the coupled oscillation of $\bm M$ and $\bm P$ (Note that the response time of $\bm M$ ($\bm P$) against the change of $\bm P$ ($\bm M$) is governed by the electron transitions among the orbital multiplets, and thus is much shorter than the typical time scale of the oscillations). To see this, we calculate the following dynamical susceptibilities by numerically solving the Landau-Lifshitz-Gilbert (LLG) equation using the fourth-order Runge-Kutta method:
%%%%%%%%%%%%%%%%%%%%%%%%%%%%%%%%%%
\begin{eqnarray}
\chi^{\rm mm}_{\alpha \beta}(\omega) &=&
\frac{\Delta M_{\alpha}^{\omega}}{\mu_0 \Delta H_{\beta}^{\omega}}
\;\;\; {\rm magnetic\;\;susceptibility,} \nonumber \\
\chi^{\rm ee}_{\alpha \beta}(\omega) &=&
\frac{\Delta P_{\alpha}^{\omega}}{\epsilon_0 \Delta E_{\beta}^{\omega}}
\;\;\; {\rm dielectric\;\;susceptibility,} \nonumber \\
\chi^{\rm em}_{\alpha \beta}(\omega) &=&
\frac{\Delta P_{\alpha}^{\omega}}
{\sqrt{\epsilon_0 \mu_0}\Delta H_{\beta}^{\omega}}
\;\;\; {\rm magnetoelectric\;\;susceptibility,} \nonumber \\
\chi^{\rm me}_{\alpha \beta}(\omega) &=&
\sqrt{\frac{\mu_0}{\epsilon_0}}
\frac{\Delta M_{\alpha}^{\omega}}{\Delta E_{\beta}^{\omega}}
\;\;\; {\rm electromagnetic\;\;susceptibility.}
\nonumber 
\end{eqnarray}
%%%%%%%%%%%%%%%%%%%%%%%%%%%%%%%%%%%
Because of the symmetry, the relation $\chi^{\rm em}_{\alpha \beta}(\omega)=\chi^{\rm me}_{\beta \alpha}(\omega)$ holds. The LLG equation is given by
%%%%%%%%%%%%%%%%%%%%%%%%%%%%%%%%%%%
\begin{equation}
\frac{d\bm m_i}{dt}=-\bm m_i \times \bm H^{\rm eff}_i 
+\frac{\alpha_{\rm G}}{m} \bm m_i \times \frac{d\bm m_i}{dt},
\label{eq:LLGEQ}
\end{equation} 
%%%%%%%%%%%%%%%%%%%%%%%%%%%%%%%%%%%
where $\alpha_{\rm G}$(=0.04) is the Gilbert-damping coefficient. The effective field $\bm H_i^{\rm eff}$ is calculated from the Hamiltonian $\mathcal{H}$=$\mathcal{H}_0$+$\mathcal{H}'(t)$ as
%%%%%%%%%%%%%%%%%%%%%%%%%%%%%%%%%%%
%%\begin{equation}
$\bm H^{\rm eff}_i = -\partial \mathcal{H} / \partial \bm m_i$.
%%\label{eq:EFFMF}
%%\end{equation}
%%%%%%%%%%%%%%%%%%%%%%%%%%%%%%%%%%%
The first term $\mathcal{H}_0$ is the model Hamiltonian given by Eq.~(\ref{eqn:model}). The perturbation term $\mathcal{H}'(t)$ represents a short rectangular pulse of magnetic field $\Delta \bm H(t)$ or electric field $\Delta \bm E(t)$, which are given, respectively, by, 
%%%%%%%%%%%%%%%%%%%%%%%%%%%%%%%%%%%
\begin{eqnarray}
\mathcal{H}^{\prime}(t)=-g\mu_{\rm B} \mu_0 \sum_i \Delta \bm H(t) \cdot \bm m_i 
\label{eqn:Hpulse}
\end{eqnarray}
%%%%%%%%%%%%%%%%%%%%%%%%%%%%%%%%%%%
and
%%%%%%%%%%%%%%%%%%%%%%%%%%%%%%%%%%%
\begin{eqnarray}
\mathcal{H}^{\prime}(t)=-\sum_i \Delta \bm E(t) \cdot \bm p_i.
\label{eqn:Epulse}
\end{eqnarray}
%%%%%%%%%%%%%%%%%%%%%%%%%%%%%%%%%%%
After applying the pulse at $t$=0, we trace time profiles of $\bm M(t)$ and $\bm P(t)$, and obtain their Fourier transforms $\Delta M_{\alpha}^{\omega}$ and $\Delta P_{\alpha}^{\omega}$. Dividing these quantities by Fourier components of the field pulses, $\Delta \bm H^{\omega}$ and $\Delta \bm E^{\omega}$, we obtain the susceptibilities. The calculations are performed using a system of $N$=20$\times$20$\times$140 sites with the periodic boundary condition.

%%%%%%%%%%%%%%%%%%%%%%%%%%%%%%%%%%%%%%%%%%%%%%%%%%%%%%%%%%%%
\begin{figure}
\includegraphics[scale=1.0]{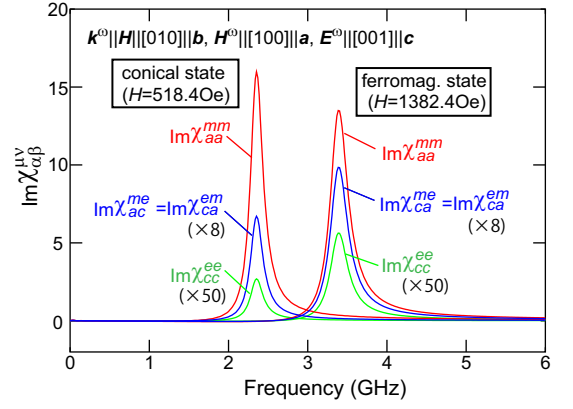}
\caption{(color online). Imaginary parts of the calculated dynamical magnetic, dielectric, and magnetoelectric susceptibilities, Im$\chi^{\rm mm}_{aa}$, Im$\chi^{\rm ee}_{cc}$, and Im$\chi^{\rm em}_{ca}$, as functions of the frequency for the conical state at $H$=518.4 Oe and the ferromagnetic state at $H$=1382.4 Oe when the static magnetic field $\bm H$$\parallel$[010]($\parallel$$\bm b$) is applied.}
%%All of these dynamical susceptibilities in each phase have peaks at the 
%%same frequency indicating the simultaneous magnetic and electric activities 
%%of the resonant modes.
\label{Fig3}
\end{figure}
%%%%%%%%%%%%%%%%%%%%%%%%%%%%%%%%%%%%%%%%%%%%%%%%%%%%%%%%%%%%%
In Fig~\ref{Fig3}, we display imaginary parts of the calculated dynamical magnetic, dielectric, and magnetoelectric susceptibilities, i.e., Im$\chi^{\rm mm}_{aa}$, Im$\chi^{\rm ee}_{cc}$, and Im$\chi^{\rm em}_{ca}$, as functions of the frequency for the conical state and the ferromagnetic state when the static magnetic field $\bm H$$\parallel$[010]($\parallel$$\bm b$) is applied. These susceptibilities have resonant peaks in the GHz regime in agreement with the microwave experiments~\cite{Onose12,Okamura13}. Moreover all of these dynamical susceptibilities in each phase have peaks at the same frequency indicating that the resonant modes have simultaneous activity to $\bm H^\omega$$\parallel$[100]($\parallel$$\bm a$) and $\bm E^\omega$$\parallel$[001]($\parallel$$\bm c$).

This magnetoelectric activity causes the microwave magnetochiral effect in Cu$_2$OSeO$_3$ [see Fig.~\ref{Fig2}(d)]. For the electromagnetic wave, the relation $\bm H^{\omega}$$\parallel$$\bm k^\omega$$\times$$\bm E^{\omega}$ holds, indicating that relative directions of the $\bm H^{\omega}$ and $\bm E^{\omega}$ components are determined by the propagation vector $\bm k^\omega$, and their relationship should be reversed upon the sign reversal of $\bm k^\omega$. When the lineally polarized electromagnetic wave with $\bm H^\omega$$\parallel$$\bm a$ and $\bm E^\omega$$\parallel$$\bm c$ propagates parallel (antiparallel) to $\bm H$$\parallel$$\bm b$, that is, sgn[Re$k^\omega$]=$+1$ (sgn[Re$k^\omega$]=$-1$) with $\bm k^\omega$=$k^\omega \hat{\bm b}$, the $\bm H^\omega$ and $\bm E^\omega$ components contribute in an additive (a subtractive) way to excite the coupled oscillation of $\bm M$ and $\bm P$, resulting in weaker (stronger) absorption of the electromagnetic wave.

%%%%%%%%%%%%%%%%%%%%%%%%%%%%%%%%%%%%%%%%%%%%%%%%%%%%%%%%%%%%%
\begin{figure}[t]
\includegraphics[scale=1.0]{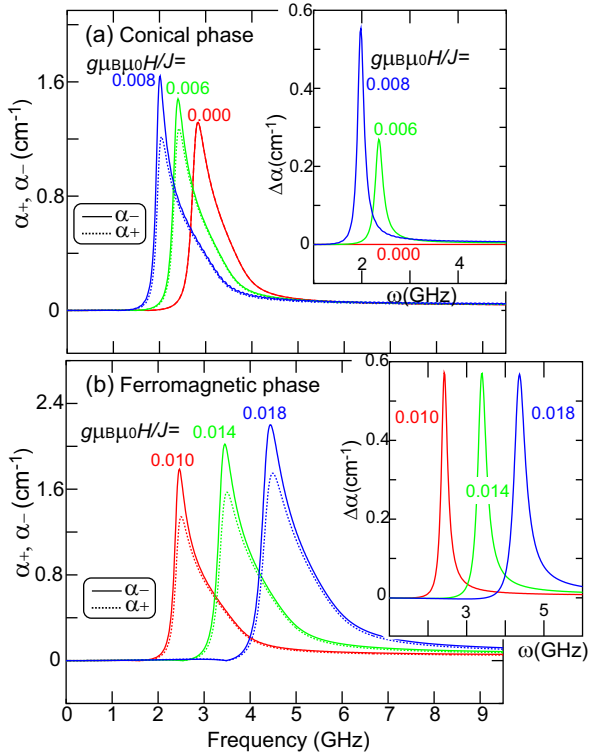}
\caption{(color online). Calculated frequency-dependence of the absorption coefficients, $\alpha_{+}$ and $\alpha_{-}$, for microwaves with sgn(Re$k^\omega$)=$+1$ and sgn(Re$k^\omega$)=$-1$, respectively, at several values of $H$ when $\bm K^{\omega}$$\parallel$$\bm H$$\parallel$$[010]$, $\bm H^{\omega}$$\parallel$$[100]$ and $\bm E^{\omega}$$\parallel$$[001]$ in the conical state (a) and the ferromagnetic state (b).}
\label{Fig4}
\end{figure}
%%%%%%%%%%%%%%%%%%%%%%%%%%%%%%%%%%%%%%%%%%%%%%%%%%%%%%%%%%%%%
%%%%%%%%%%%%%%%%%%%%%%%%%%%%%%%%%%%%%%%%%%%%%%%%%%%%%%%%%%%%%
\begin{figure}[t]
\includegraphics[scale=1.0]{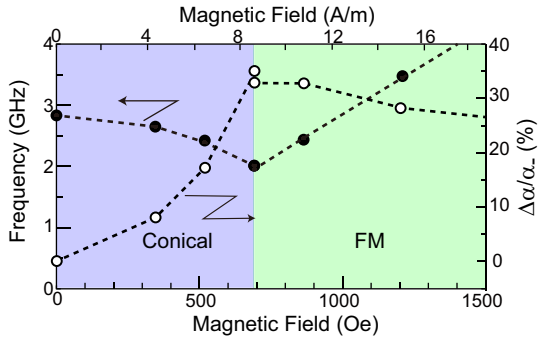}
\caption{(color online). Calculated magnetic-field dependence of resonance frequency and magnitude of the magnetochiral dichroism $\Delta \alpha/\alpha_-$.}
\label{Fig5}
\end{figure}
%%%%%%%%%%%%%%%%%%%%%%%%%%%%%%%%%%%%%%%%%%%%%%%%%%%%%%%%%%%%%
The expression of the complex refractive index $N(\omega)$ is derived by solving the Fourier-formed Maxwell's equations as~\cite{Kezsmarki11},
%%%%%%%%%%%%%%%%%%%%%%%%%%%%%%%%%%%
\begin{eqnarray}
N(\omega) =\frac{c}{\omega}k^\omega &\sim&  \sqrt{
[{\epsilon_{cc}}^\infty + \chi^{\rm ee}_{cc} (\omega)]
[{\mu_{aa}}^\infty + \chi^{\rm mm}_{aa}  (\omega)]}
\nonumber \\
&+& sgn({\rm Re}k^\omega) [\chi^{{\rm me}}_{ac} (\omega)
+ \chi^{{\rm em}}_{ca} (\omega)]/2,
\label{eqn:CRI1}
\end{eqnarray}
%%%%%%%%%%%%%%%%%%%%%%%%%%%%%%%%%%%
for $\bm k^\omega$=$k^\omega \hat{\bm b}$$\parallel$[010], $\bm H^\omega$$\parallel$[100]$\parallel$$\bm a$ and $\bm E^\omega$$\parallel$[001]$\parallel$$\bm c$. The absorption coefficient $\alpha(\omega)$ is related to $N(\omega)$ as,
%%%%%%%%%%%%%%%%%%%%%%%%%%%%%%%%%%%
\begin{eqnarray}
\alpha(\omega)=\frac{2\omega\kappa(\omega)}{c} 
=\frac{2\omega}{c} {\rm Im}N(\omega),
\label{eqn:alpha}
\end{eqnarray}
%%%%%%%%%%%%%%%%%%%%%%%%%%%%%%%%%%%
and thus attains the directional dependence via the sign of Re$k^\omega$.
Here $\kappa(\omega)={\rm Im}N(\omega)$ is the extinction coefficient.

Figures~\ref{Fig4}(a) and (b) display calculated frequency-dependence of $\alpha_+$ and $\alpha_-$ in the conical phase and the ferromagnetic phase, respectively, for several values of $H$, where $\alpha_+$ and $\alpha_-$ are the absorption coefficients for microwaves propagating parallel and antiparallel to $\bm H$$\parallel$[010], respectively. In the calculation, we take $\epsilon_{zz}^{\infty}$=8 and $\mu_{zz}^{\infty}$=1 according to the experimental data~\cite{Belesi12,Miller10}. We find that the microwave absorption is resonantly enhanced at the eigen-frequency of the electromagnon excitation, and there exists significant difference between $\alpha_+$ and $\alpha_-$.

Calculated magnitude of the directional dichroism $\Delta\alpha$/$\alpha_-$ where $\Delta\alpha=\alpha_+ - \alpha_-$ is plotted in Fig.~\ref{Fig5}. This quantity is governed by the amplitude of ferroelectric polarization $P$, which is nearly proportional to the square of the net magnetization $M^2$. The dichroism increases in the conical state with the growth of $P$ and $M$ as the magnetic field increases. In contrast, once the system enters the ferromagnetic state, the saturated $M$ gives nearly constant directional dichroism although there still exists slight field-dependence due to the field-dependent resonant frequency via $\omega$ in Eq.~(\ref{eqn:alpha}). The directional dichroism is enhanced at the phase boundary between the conical and the ferromagnetic phases, and its magnitude reaches as much as $\sim$30$\%$. Calculated electromagnon resonance frequency plotted in Fig.~\ref{Fig5} shows decreasing behavior in the conical state, while increasing behavior in the ferromagnetic state as the magnetic field increases. This behavior is in good agreement with the experimental observations~\cite{Onose12,Okamura13} and coincides with analytical formula of the spin-wave gap given in Ref.~\cite{Kataoka87}.

In summary, we have theoretically predicted that magnetically ordered phases in the chiral multiferroics Cu$_2$OSeO$_3$ host gigantic microwave magnetochiral dichroism. This phenomenon results from interference between the magnetic and electric activation processes of electromagnons with GHz resonance frequencies. It has been demonstrated that long-period magnetic structures in the chiral multiferroics without inversion symmetry can host gigantic dynamical magnetoelectric phenomena at GHz regime.
%%and thus provide useful microwave-device functions.
In order to further enhance the effect, search for novel chiral multiferroics with larger $\bm P$ or stronger magnetoelectric coupling is needed. For example, chiral multiferroics based on the inverse Dzyaloshinskii-Moriya mechanism~\cite{Katsura05} as an origin of its $\bm P$ is worth trying to search because this mechanism tends to induce large $\bm P$ relative to the spin-dependent metal-ligand hybridization mechanism in Cu$_2$OSeO$_3$. Metamaterials, thin films, and synthetic nanomaterials are another promising candidate to realize enhanced microwave ME effects through artificially designing large magnetoelectric susceptibilities and intense electromagnon resonances~\cite{Sihvola07,Urzhumov08,Zivieri12,Vlaminck08,Vlaminck10,Bilzer07}.

The author thanks I. K\'ezsm\'arki, F. Kagawa, Y. Okamura, S. Seki, Y. Tokura for discussions. This research was in part supported by JSPS KAKENHI (Grant Numbers 25870169 and 25287088).
%%%%%%

\end{document}